# Hardwire-Configurable Photonic Integrated Circuits Enabled by 3D Nano-Printing

Tobias Hoose,[1,2,†] Matthias Blaicher,[1,2,†] Juned Nassir Kemal,[2]
Heiner Zwickel,[2] Muhammad Rodlin Billah,[1,2,3]
Philipp-Immanuel Dietrich,[1,2,3] Andreas Hofmann,[4] Wolfgang Freude,[2]
Sebastian Randel[2] and Christian Koos[1,2,3*]

[1]Institute of Microstructure Technology (IMT), Karlsruhe Institute of Technology (KIT), Hermann-von-Helmholtz-Platz 1, 76344 Eggenstein-Leopoldshafen, Germany
[2]Institute of Photonics and Quantum Electronics (IPQ), KIT, Engesserstrasse 5, 76131 Karlsruhe, Germany
[3]Vanguard Automation GmbH, Gablonzer Strasse 10, 76185 Karlsruhe, Germany
[4]Institute of Applied Computer Science (IAI) KIT, 76344 Eggenstein-Leopoldshafen, Germany
[†]Now with Nanoscribe GmbH, Hermann-von-Helmholtz-Platz 1, 76344 Eggenstein-Leopoldshafen, Germany
*Corresponding author: christian.koos@kit.edu

**Abstract:** Photonic integrated circuits (PIC) currently rely on application-specific designs that are geared towards a particular functionality. Development of such application-specific PIC requires considerable effort and often involves several design, fabrication, and characterization iterations, each with cycle times of several months, to reach acceptable performance. At the same time, the number of chips that is eventually needed to serve a certain application is often too small to benefit from the scalability of advanced photonic integration platforms. As a consequence, large-scale photonic integration has hitherto been unable to unfold its full economic advantages within the highly fragmented application space of photonics. Here we introduce a novel approach to configurable PIC that can overcome these challenges. The concept exploits generic PIC designs consisting of standard building blocks that can be concatenated to provide an application-specific functionality. Configuration of the PIC relies on establishing hardwired optical connections between suitable ports of an optical wire board using highly flexible 3D direct-write laser lithography. Hardwire-configurable PIC allow to exploit high-throughput production of generic chips in large-scale fabrication facilities for serving a wide variety of applications with small or medium volumes of specifically configured devices.

## 1. INTRODUCTION

Photonic integration is a key technology that can serve a wide variety of applications, ranging from high-speed communications [1-3], optical signal processing [4,5] and microwave photonics [6-8] to high-precision metrology [9,10] and optical sensing [11-13]. Over the last years, the technology has experienced tremendous progress, and several platforms have reached industrial maturity [1, 14-18]. The most prominent example is silicon photonics (SiP), which exploits highly developed CMOS processes for low-cost mass-production of PIC [1,2,13,15,17,21]. For these circuits, design and mask fabrication represent the main cost drivers, whereas large-scale mass production of a specific PIC design is available at rather low cost through foundry-based fabrication services worldwide [15,17,18,22]. However, this situation is somewhat in contrast to the fact that photonics still represents a highly fragmented market, covering a wide variety of applications that require low or medium quantities of application-specific PIC rather than large quantities of the same PIC [17,18]. As a consequence, highly scalable photonic integration can currently not unfold its full economic strengths.

This contradiction can be overcome by circuit architectures that contain a multitude of generic building blocks and that can be programmed to provide specific functionalities in a separate configuration step. Configurable PIC have been previously demonstrated, exploiting, e.g., meshes of power splitters and thermo-optic phase shifters that can be tuned to control the flow of light through the network and thus provide reconfigurability [4,5,23-25]. However, while these concepts have led to impressive demonstrations of reconfigurable optical filters [4,23-25], they rely on precise tuning and permanent tracking of a large number of electrical signals to address the various phase shifters and to maintain a certain functionality. This is associated with additional control overhead and significant power dissipation, in particular when it comes to large-scale circuits. Moreover, optical crosstalk in the various switching elements of mesh-based photonic signal processors may add up and hence limit the scalability of the concept.

In this paper we introduce and experimentally demonstrate the concept of hardwire-configurable PIC (HC-PIC) that allows to flexibly combine a multitude of standard building blocks designed, e.g., for light manipulation, routing, and detection. The application-specific functionality of HC-PIC is established by hardwired concatenation of functional building blocks in a one-time configuration step. While manufacturing of the underlying generic PIC can rely on high-throughput production of standardized designs in large-scale fabrication facilities, the application-specific circuit configuration is established during the packaging process by connecting suitable ports of an optical wire board (OWB) using highly flexible 3D direct-write laser lithography [26-28]. By appropriate choice of on-chip building blocks, the same generic PIC can be used to serve a wide variety of applications. In our experiments, we demonstrate the viability of the HC-PIC concept by using a generic circuit design that can be configured to act as transmitter and/or as a receiver for different kinds of data signals. To the best of our knowledge, this is the first demonstration of hardwire-configurability in photonic integration.

## 2. FABRICATION AND CONCEPT OF HC-PIC

The fabrication and configuration workflow of the HC-PIC is illustrated in Fig. 1(a). Established wafer-scale processes are used for efficient fabrication of identical PIC in large quantities. These PIC are generic in the sense that they carry a multitude of standard building blocks offering a variety of different functionalities. The application-specific functionality circuit is established by appropriate concatenation of such building blocks in a dedicated configuration step, which exploits highly flexible direct-write laser lithography for in-situ printing of 3D freeform single-mode waveguides that connect suitable ports of an OWB. The concept allows using the same generic PIC design for serving a wide variety of applications, ranging from optical communications and signal processing to optical metrology and sensing.

The schematic concept of the HC-PIC, designed for our experiments is shown in Fig. 1(b). The PIC was fabricated through a commercially available foundry service [29] and relies on external light sources. The circuit comprises a variety of passive elements such as power splitters and combiners, coupling elements for off-chip connections, phase shifters and electro-optic modulators for light manipulation, as well as photodetectors for light detection. All these building blocks are connected to the OWB, which allows free configuration of the PIC functionality by establishing appropriate connections between the various optical ports. The OWB contains a multitude of coupling sites realized by tapered silicon nanowire waveguides, onto which the OWB connections can be printed, see Supplementary Section S1 and S2 for details.

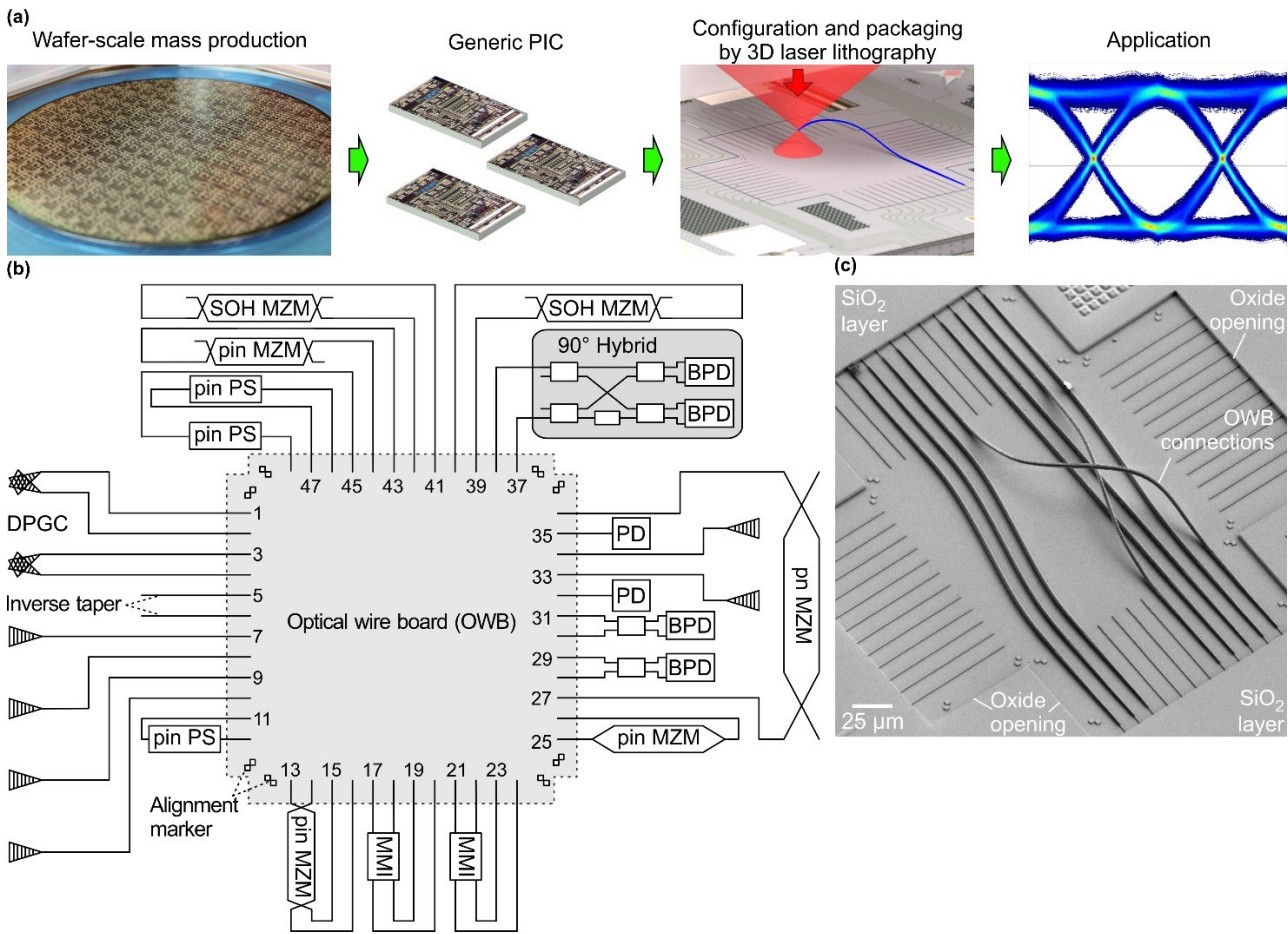

**Fig. 1.** Concept of hardwire-configurable photonic integrated circuits (HC-PIC). (a) Fabrication and configuration workflow of HC-PIC: Established wafer-scale processes allow for efficient fabrication of identical generic PIC in large quantities. The generic PIC carries a multitude of standard building blocks for, e.g., light generation, manipulation, and detection. The application-specific functionality of the PIC is established by appropriately connecting the available building blocks in a one-time configuration step. The connections are created by direct-write laser lithography, i.e., by in-situ printing of 3D freeform single-mode waveguides. The procedure is similar to photonic wire bonding, which used for connecting the HC-PIC to optical fibres [30] and other photonic dies [27] (not shown). Configuration and packaging of the HC-PIC can hence be combined in a common fabrication step. HC-PIC allow to use a single generic design to address a wide variety of applications, ranging from optical communications and signal processing to optical metrology and sensing. (b) Layout of the silicon PIC used for subsequent experiments. The PIC comprises a variety of devices: Conventional grating couplers (GC), dual-polarization grating couplers (DPGC), and inversely tapered silicon waveguides for coupling of light to and from the chip, optical p-i-n-type phase shifters (pin PS) as well as silicon-organic hybrid (SOH), p-i-n-type, and depletion-type p-n-Mach-Zehnder modulators (SOH MZM, pin MZM, pn MZM), multi-mode interference couplers (MMI) acting as power splitters and combiners, and GeSi photodetectors (PD), both as single devices and as balanced photodetectors (BPD). All these blocks are accessible via the ports 1…48 of the optical wire board (OWB). (c) Scanning-electron microscope (SEM) image of a PIC with a hardwire-configured OWB. The chip surface is covered with a protective oxide, which is locally opened for accessing the OWB ports.

Free configurability of the connections requires in-situ fabrication of compact waveguides that allow low-loss and low crosstalk single-mode transmission. This is achieved by using high-resolution two-photon lithography to create the 3D freeform waveguide structures. After lithography and development, the free-standing waveguides are embedded into a low-refractive-index material that acts both as a waveguide cladding and as a protective cover. Note that wiring of the OWB relies on the same tools and processes as the concept of photonic wire bonding (PWB), which can be used for connecting HC-PIC to optical fibres [30] and external light sources [10] or for realizing advanced multi-chip modules [31] crossing-free PIC [31]. Packaging and configuration of the HC-PIC can hence be efficiently combined in a single fabrication step. Fig. 1(c) shows a scanning-electron microscope (SEM) image of a configured OWB. The OWB contains an overall number of 48 ports, 16 of which are connected in the illustrated configuration. Neighbouring ports are spaced by 10 µm, thus permitting high wiring density. The OWB connections have a rectangular cross-section of 2.0 µm × 1.8 µm and are fabricated using a commercially available photoresist with a refractive index of $n_{core}$ = 1.53. This allows single-mode operation after cladding with low-refractive-index material of $n_{clad}$ = 1.36. Details of the OWB design and the fabrication of the hardwired connections can be found in Supplementary Section S1.

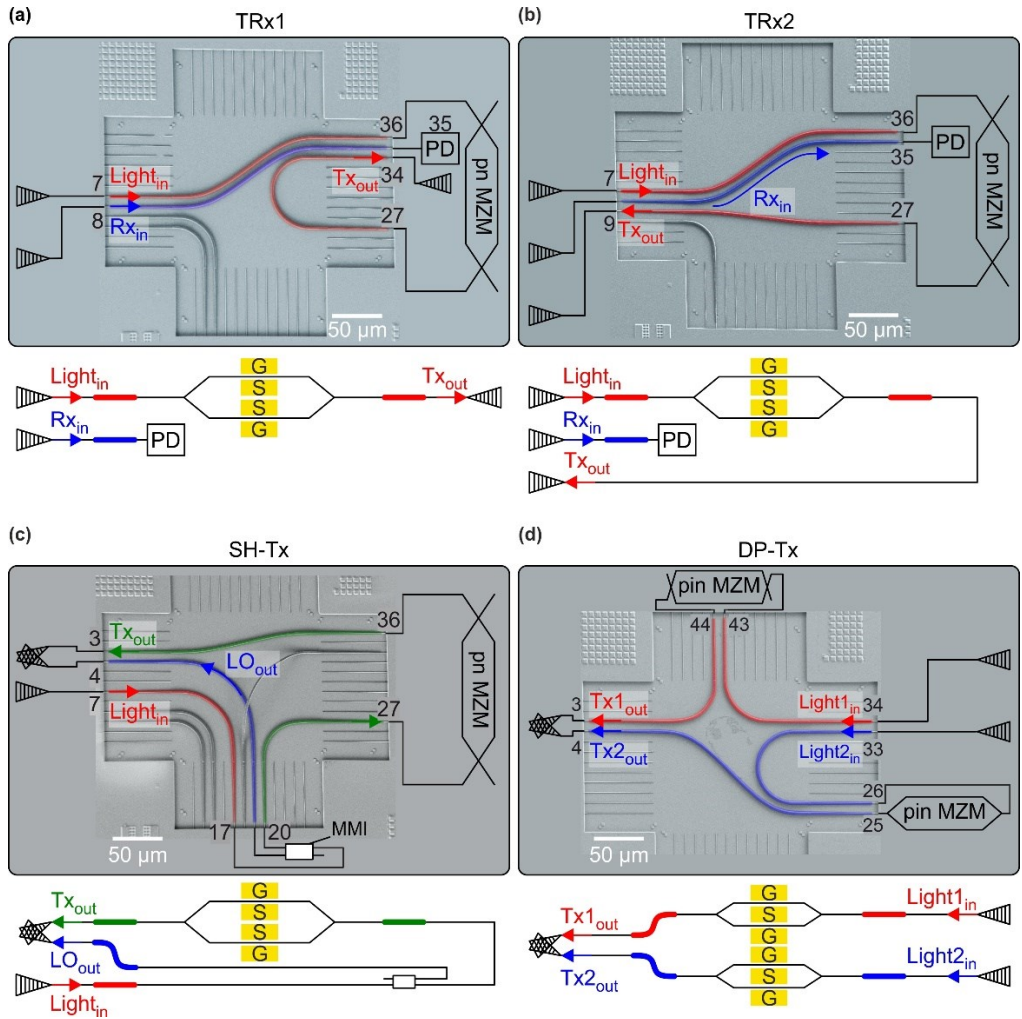

**Fig. 2.** Implementation of four different hardwire-configurable photonic integrated circuits (HC-PIC). (a) First implementation of a single-channel optical transceiver (TRx1) for intensity modulation and direct detection (IM/DD). The circuit comprises a transmitter (red OWB connections) with light input port (Light$_{in}$) and output port (Tx$_{out}$) and a receiver (blue OWB connection) with input Rx$_{in}$. All fiber-chip connections are realized by grating couplers (GC). The transmitter input port and the output port are on opposite sides of the die for facilitating coupling with GC and standard single mode fibres (SMF). (b) Second implementation of a single-channel optical transceiver (TRx2), configured for coupling to all optical ports through a single fibre array with a fiber-to-fiber pitch of 127 µm. OWB connections of the transmitter are again depicted in red, whereas the blue connection belongs to the Rx. (c) Self-homodyne transmitter (SH-Tx), configured for transmitting a data signal (Tx$_{out}$) and an unmodulated local-oscillator tone (LO$_{out}$) on orthogonal polarization states of a SMF. Green OWB connections belong to the transmitter for the data signal, the blue connection is part of the LO transmitter, and the red connection is used for transport of input light from an external source. For exploring the full benefits of coherent self-homodyne transmission, the MZM should be replaced by an integrated IQ modulator. (d) Dual-polarization transmitter (DP-Tx). Two independent data streams are generated by a pair of MZM and then coupled to two orthogonal polarization states of a SMF through a dual-polarization grating coupler [33] (DPGC). OWB connections belonging to the two transmitters are depicted in blue and red. In a slightly different OWB configuration, a single optical input port might be used to feed both modulators.

## 3. IMPLEMENTATIONS AND CHARACTERIZATION

To demonstrate the viability of our concept, we use a generic HC-PIC to implement four different photonic devices for optical communications, comprising two transceivers (TRx) for data transmission via intensity modulation and direct detection (IM-DD) as well as a self-homodyne and a dual-polarization transmitter, see Fig. 2(a)-(d) for schematics and SEM images of the associated OWB configurations.

### A. Single-channel transceivers

For the two single-channel TRx, we use the OWB connections coloured in red for the transmitter part and blue for the receiver, see Fig. 2(a) and (b). For both versions, the transmitter consists of a depletion-type SiP Mach-Zehnder Modulator (pn MZM) with its input and output waveguides (Port 36 and Port 27) connected to grating couplers (GC) via the OWB. For Transceiver 1 (TRx1), Fig. 2(a), the modulator is connected to GC (Port 7 and Port 34) of the HC-PIC, which are placed at opposite edges of the chip to enable in and out coupling with two independent standard single mode fibres (SMF). The overall insertion loss (IL) of the concatenated OWB connections (7−36) and (27−34) (red) amounts up to 4.0 dB, obtained by first measuring the transmission of the whole device and by taking into account GC and modulator losses that were obtained from reference measurements, see Supplementary Section S3 for details. Note that for this configuration of OWB connections, separation of the individual IL contributions is not possible. The IL of the OWB connection to the on-chip photodiode (8−35) of the receiver is 0.9 dB. Transceiver 2 (TRx2), Fig. 2(b), is configured such that grating couplers of all input and output ports are placed on the same side of the chip for coupling to a fibre array. In this case, the overall IL of the concatenated OWB connections (7−36) and (27−9) amounts again to 4.0 dB, whereas the IL of the OWB connection to the receiver (8−35) is found to be 1.8 dB. We attribute the differences to non-ideal OWB coupling interfaces, caused by slightly over-etched inversely tapered SiP waveguides, see Supplementary Section S2 for details. We expect that the IL of the OWB connections can be significantly reduced by improved design and fabrication of the OWB, allowing, e.g., for bigger bending radii of the waveguide trajectories.

### B. Self-homodyne transmitter

The implementation of the self-homodyne transmitter is shown in Fig. 2(c). The device configuration allows transmission of a data signal and a continuous-wave (CW) local-oscillator (LO) tone on two orthogonal polarization states of the same SMF. Deriving the signal carrier and the LO tone from the same laser allows for homodyne detection with small phase noise and greatly simplifies the RX architecture [33]. The transmitter consists of a GC for CW light input at Port 7, a multimode-interference coupler (MMI) that splits the incoming CW light (Port 17) into the LO tone (Port 19) and the signal carrier (Port 20), a depletion-type SiP MZM (input at Port 27, output at Port 36), and a polarization-diversity grating coupler [34] (DPGC, Port 3 and 4) for coupling the light to orthogonal polarization states of the SMF. Note that the current implementation of the self-homodyne transmitter as shown in Fig. 2(c) is of rather limited practical use since the MZM allows generation of amplitude shift-keying signals only. To exploit the full potential of self-homodyne transmission, higher-order modulation formats such as quadrature amplitude modulation (QAM) would be preferred, which would require an IQ modulator rather than a simple MZM. For the implementation shown in Fig. 2(c), an overall IL of 3.3 dB is measured for the concatenated OWB connections (7−17) and (19−4) along the LO path. The signal path also contains the OWB connection (7−17), which is now concatenated with connections (20−27) and (36−3), leading to an overall IL of about 7.6 dB. Also here, the IL is mainly caused by imperfect coupling interfaces, see Supplementary Section S2 and S3 for details. Note that this HC-PIC would also allow realization of a self-homodyne receiver, see Supplementary Section S5 for details.

### C. Dual-polarization transmitter

For the dual-polarization transmitter, Fig. 2(d), two independent data signals are generated by a pair of Mach-Zehnder modulators and subsequently coupled to orthogonal polarization states of an SMF. In this device, an overall IL of 1.7 dB is measured for the concatenated OWB connections (34−43) and (44−3) along the first transmitter path. For the second transmitter path, the overall IL of the concatenated OWB connections (33−26) and (25−4) amounts to 3.1 dB. Note that, by including an additional power splitter, the dual-polarization transmitter can also be configured to distribute light from a single optical source to both modulators, see Supplementary Section S5 for details. Also here, the MZM can be replaced by IQ modulators in future implementations to fully exploit the benefits of polarization-diverse coherent transmission. Note further that our current implementation of the dual-polarization transmitter relies on rather slow injection-type MZM with p-i-n junctions acting as phase shifters in both arms. These devices are limited to modulation speeds of a few Gbit/s only, and the implementation rather serves illustrative purposes to show the flexibility of OWB concept. The devices should be replaced by faster p-n depletion-type MZM to obtain practically relevant transmitter performance.

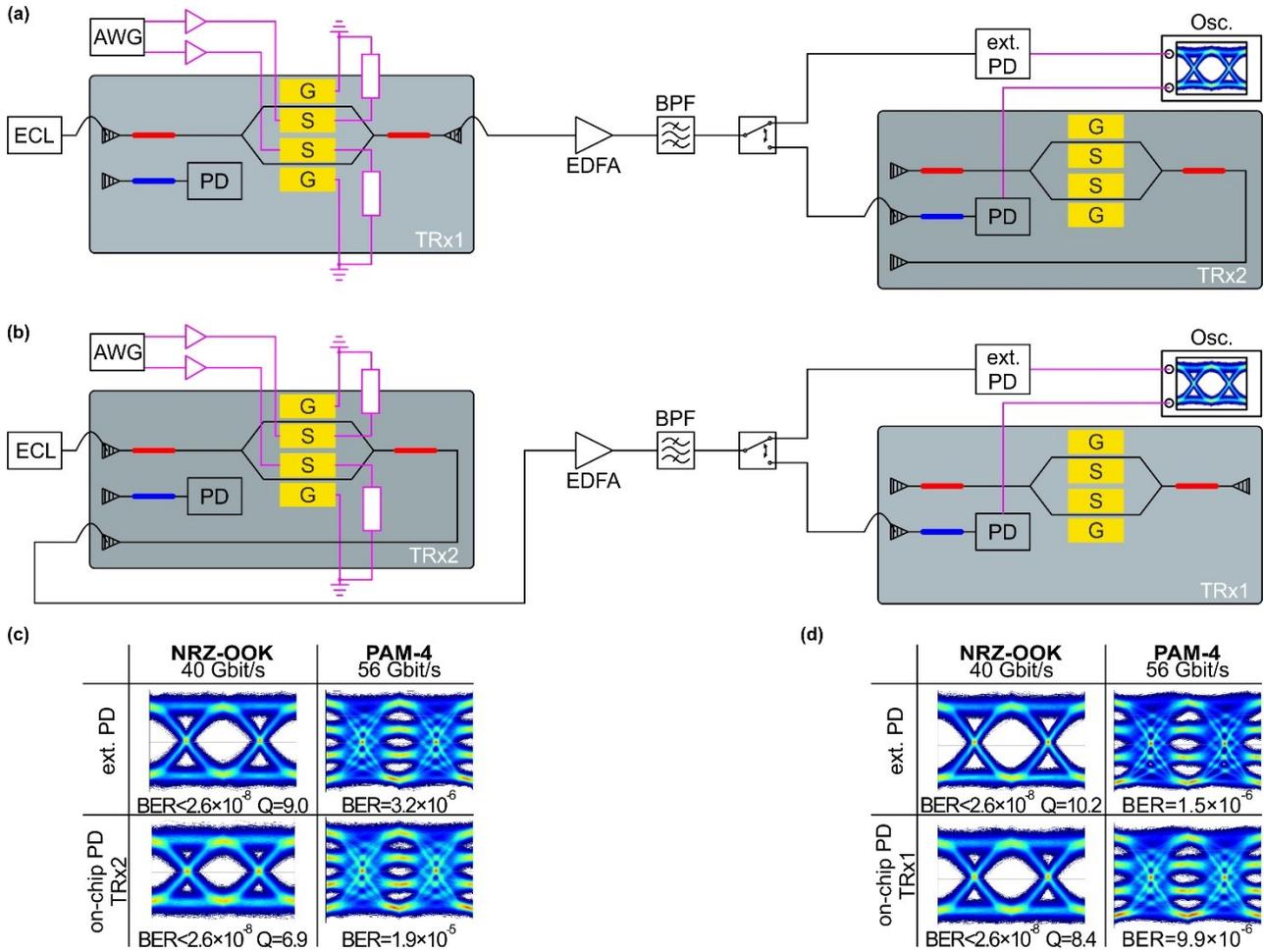

**Fig. 3.** Experimental demonstration of HC-PIC for data transmission with intensity modulation and direct detection (IM-DD). The demonstrations rely on the transceiver implementations TRx1 and TRx2 depicted in Fig. 2(a) and Fig. 2(b). (a) Transmission demonstration using TRx1 as a transmitter. The signals are detected either by an on-chip photodiode (PD) of TRx2 by an external PD (ext. PD) for benchmarking. The MZM are driven by an arbitrary-waveform generator (AWG), see Supplementary Section S4 for details. Light is supplied by a benchtop-type external-cavity laser (ECL). Erbium doped fibre amplifiers (EDFA) are used to compensate the fibre-chip coupling and propagation losses. A bandpass filter (BPF) suppresses out-of-band amplified spontaneous emission noise. The received electrical signal is recorded with a high-speed oscilloscope (Osc.) for offline processing, see Supplementary Section S4. (b) Transmission demonstration using TRx2 as a transmitter. This setup is logically identical to Fig. 2(a), except that different ports on the HC-PIC are used for in and out-coupling of light. (c) Results of the transmission experiments according to Fig. 3(a) using OOK signalling at a line rate of 40 Gbits and and PAM-4 signalling at a line rate of 56 Gbit/s. For the OOK signaling, we could not measure any error within our recording length of $3.9 \times 10^7$ bits. We can hence only specific an upper bound of $2.6 \times 10^{-8}$ for the BER. In general, the use of the on-chip PD results in slightly narrower eye openings, but the recordings do still not contain any errors for OOK and the BER is below the threshold for forward error correction with 7 % overhead in the case of PAM-4. (d) Results of the transmission experiments according to Fig. 3(b). We again use OOK and PAM-4 line rates of 40 Gbit/s and 56 Gbit/s, respectively. Also here, we could not find any errors in our OOK recordings, and the PAM-4 BER is below the threshold for forward error correction with 7 % overhead.

## 4. DATA TRANSMISSION EXPERIMENTS

### A. IM-DD transceivers

For demonstrating the functionality of the IM-DD transceivers TRx1 and TRx2, we send data from one device to the other using the setups depicted in Fig. 3(a) and (b). An arbitrary-waveform generator (AWG) and a pair of RF amplifiers are used to generate the electric drive signals, which are coupled to the MZM of TRx1 through microwave probes in ground-signal-signal-ground (G-S-S-G) configuration. The MZM is fed by an external-cavity laser (ECL) via GC, and the data signal is sent through an erbium-doped fibre amplifier (EDFA) to compensate for fibre-chip coupling losses, followed by a noise-blocking band-pass filter (BPF) for suppression of amplified spontaneous emission (ASE) noise. In a first experiment, we generate simple NRZ signals at data rates of 40 Gbit/s, and we detect them by an external photodetector, which is connected to a high-speed oscilloscope (Osc.). The results serve as a benchmark for a second experiment, in which we use the on-chip photodiode of TRx2 to receive the signal. The eye diagrams obtained from the external and the on-chip PD are shown in in the first column of Fig. 3(c). In both cases, we do not find any errors in our recordings of $3.9 \times 10^7$ bits, giving an upper bound of $2.6 \times 10^{-8}$ for the BER, well below the threshold of BER = $3.8 \times 10^{-3}$ for hard-decision forward error correction (HD-FEC) with 7 % coding overhead. From the eye diagrams, we estimate Q factors of 9.0 and 6.9 for the external photo-receiver and for the on-chip photodiode of TRx2, respectively. In both cases, this would correspond to error-free transmission with a BER of well below $10^{-9}$ assuming that the signals are impaired by additive white Gaussian noise only. We repeat the experiments using four-level pulse amplitude modulation (PAM-4) at a symbol rate of 28 GBd, leading to a data rate of 56 Gbit/s, see second column of Fig. 3(c). In these experiments, BER values of $3.2 \times 10^{-6}$ and $1.9 \times 10^{-5}$ were measured from our recordings – well below the 7 % HD FEC limit. In all demonstrations, we used both pre- and post-equalization to compensate for the channel frequency response, see Supplementary Section S4 for details. The same set of experiments is performed with TRx2 acting as a transmitter and TRx1 as a receiver, see Fig. 3(b) for the experimental setup and Fig. 3(d) for the resulting eye diagrams. As before, all measured BER remain below the threshold for 7 % HD FEC, while the OOK experiments may even be assumed to be error-free based on the measured Q-factors.

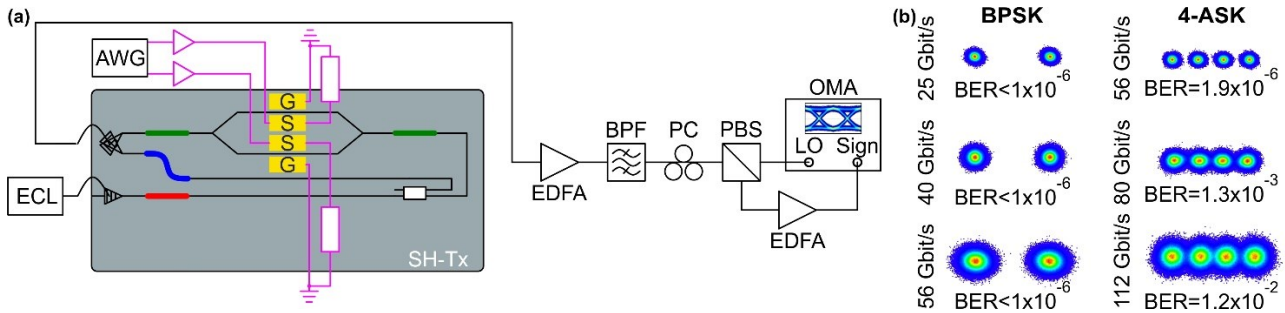

**Fig. 4.** Experimental demonstration of self-homodyne transmitter (SH-Tx) realized as HC-PIC. (a) Experimental setup. The demonstration relies on the transmitter implementation shown in Fig. 2(c). Light is supplied by an external-cavity laser (ECL). The data signal and the unmodulated LO tone are coupled to orthogonal polarization states of the transmission fibre. After an EDFA and a bandpass filter (BPF), the polarization of the overall signal is adjusted by a polarization controller (PC) such that data signal and the LO tone can be separated by a simple polarization beam splitter (PBS). The signal and the LO tone are then fed to an optical modulation analyzer (OMA) for extracting the data and for characterizing the signals. (b) Constellation diagrams for binary phase shift keying (BPSK) at line rates of 25, 40, and 56 Gbit/s, and for four-level amplitude shift keying (4-ASK) at 56, 80, and 112 Gbit/s. For the BSPK experiments, we could not find any errors within our evaluated signals each of which comprises $10^6$ bits, leading to an upper limit of $1 \times 10^{-6}$ for the BER. This is well below the limit of $3.8 \times 10^{-3}$ for FEC with 7 % overhead. The same is true for a ASK transmission at 56 Gbit/s and 80 Gbit/s. For 4-ASK at 112 Gbit/s, soft-decision FEC with a slightly overhead of 20 % would be needed.

## B. Self-homodyne transmitter experiment

For demonstration of the self-homodyne transmitter shown Fig. 2(c), we use the setup depicted in Fig. 4(a). We transmit BPSK signals at data rates of up 56 Gbit/s and four-level amplitude-shift-keying (4-ASK) signals with symbol rates of up to 56 Gbd, corresponding to a data rate of 112 Gbit/s. The MZM is again contacted by microwave probes in G S S G configuration and driven by RF signals generated by an AWG. The transmitted signals are amplified by an EDFA and sent through a bandpass filter for ASE noise suppression. The receiver is realized from discrete elements and comprises a polarization beam splitter (PBS) that separates the unmodulated LO tone from the data signal before feeding them to an optical modulation analyzer (OMA). Signal processing is performed offline, see Supplementary Section S4. The corresponding constellation diagrams and the associated BER values are depicted in Fig. 4(b). Note that this experiment serves rather illustrative purposes in future implementations, the MZM should be replaced by an IQ modulator to exploit the full potential of coherent self-homodyne transmission. Still, the results illustrate the potential of HC PIC to serve a wide variety of applications using a single generic PIC layout. Note that the application potential of HC-PIC is not limited to optical transmitters and receivers, but also covers various implementations of sensing and measurement systems. As an example, appropriate concatenation of the building blocks shown in Fig. 1(b) would allow to realize integrated optical chip-scale optical coherence tomography (OCT) receivers [12] or integrated optical circuits for high-precision dual-comb distance metrology [9].

## 5. SUMMARY

In summary, we have introduced and experimentally demonstrated the concept of hardwire configurable photonic integrated circuits (HC PIC), which exploits highly flexible 3D direct write laser lithography to define application-specific functionalities of a photonic die during the packaging process. In our experiments, we use a generic PIC and realize four optical transmitter and receiver implementations by appropriate concatenation of on-chip building blocks via an optical wire board (OWB). We achieve line rates of more than 100 Gbit/s per wavelength with significant potential for further improvement by optimized design of the on-chip devices and by an improved implementation of the OWB. The concept allows to exploit highly standardized mass production of silicon photonic (SiP) circuits to serve a highly fragmented application space with small or medium quantities of specifically configured devices.

*See Supplementary below for supporting content.*

**Funding.** This work was supported by the Cluster of Excellence "3D Matter Made to Order" (3DMM2O), by the German Federal Ministry of Education and Research (Bundesministerium für Bildung und Forschung, BMBF) within projects PHOIBOS (# 13N12574) and PRIMA (# 13N14630), by the H2020 Photonic Packaging Pilot Line PIXAPP (# 731954), by the European Research Council (ERC Consolidator Grant 'TeraSHAPE', # 773248), by the Deutsche Forschungsgemeinschaft (DFG) through the CRC WavePhenomena (# 1173) Project C4, by the Helmholtz International Research School for Teratronics (HIRST), and by the Alfried Krupp von Bohlen und Halbach Foundation. The chip designs in this study made use of design elements in the OpSIS O150A PDK.

**Acknowledgments.** We thank M. Lauermann for designing the HC-PIC, C. Naber for supporting the experiments, and S. Wolf for discussions about HC-PIC configurations.

**Disclosures.**  P.-I. D. and C. K. are co-founders and shareholders of Vanguard Photonics GmbH, a start-up company offering tools and processes for 3D nano printing in the field of photonic integration and assembly. P.-I. D. and M. R. B. are employees of Vanguard Automation GmbH, a subsidiary of Vanguard Photonics GmbH. M. B. is co-inventor of a patent owned by Karlsruhe Institute of Technology (KIT) in the technical field of the publication. T. H. and M. B. are now employees of Nanoscribe GmbH, a company selling 3D lithography systems.

# Hardwire-Configurable Photonic Integrated Circuits Enabled by 3D Nano-Printing

Tobias Hoose,[1,2,†] Matthias Blaicher,[1,2,†] Juned Nassir Kemal,[2]
Heiner Zwickel,[2] Muhammad Rodlin Billah,[1,2,3]
Philipp-Immanuel Dietrich,[1,2,3] Andreas Hofmann,[4] Wolfgang Freude,[2]
Sebastian Randel[2] and Christian Koos[1,2,3*]

[1]*Institute of Microstructure Technology (IMT), Karlsruhe Institute of Technology (KIT), Hermann-von-Helmholtz-Platz 1, 76344 Eggenstein-Leopoldshafen, Germany*
[2]*Institute of Photonics and Quantum Electronics (IPQ), KIT, Engesserstrasse 5, 76131 Karlsruhe, Germany*
[3]*Vanguard Automation GmbH, Gablonzer Strasse 10, 76185 Karlsruhe, Germany*
[4]*Institute of Applied Computer Science (IAI) KIT, 76344 Eggenstein-Leopoldshafen, Germany*
[†]*Now with Nanoscribe GmbH, Hermann-von-Helmholtz-Platz 1, 76344 Eggenstein-Leopoldshafen, Germany*
*\*Corresponding author: christian.koos@kit.edu*

**Abstract:** This document provides supplementary information to "Hardwire-Configurable Photonic Integrated Circuits Enabled by 3D Nano-Printing". It contains details about methods used for the fabrication of OWB connections and for the experimental demonstration, as well as details of the insertion loss measurements. Further functionalities of the generic PIC, which are mentioned in the primary document, are summarized in the last section of the supplementary material.

## S1. FABRICATION OF OWB CONNECTIONS

All OWB connections were printed using a modified commercial two-photon lithography system (Nanoscribe GmbH, Photonic Professional GT), equipped with a 40× objective lens (numerical aperture 1.4) and galvanometer mirrors, allowing for fast beam movement in the lateral direction. The lithography system is equipped with a proprietary control software that allows for precise localization of coupling interfaces as well as for fabrication of OWB connections with high shape fidelity. We use a commercially available photoresist (IP-Dip by Nanoscribe GmbH), having a refractive index of $n_{core}$ = 1.53 at a wavelength of 1.55 µm. Precise alignment of printed waveguide structures with respect to the OWB coupling sites is accomplished in a two-step process, involving camera-based detection of reference markers as well as confocal imaging of on-chip structures. After exposure, the samples are developed in propylene-glycol-methyl-ether-acetate (PGMEA) and subsequently rinsed in isopropyl alcohol (2-propanol). For optical coupling from the SiP nanowire to the printed OWB connection, we use a double taper structure, see Supplementary Section S2 and Supplementary Fig. S1 for details. In our experiments, the cladding material is mimicked by an index-matching oil (Cargille Refractive Index Liquids Cat#:18031 Series AAA) with a refractive index of $n_{clad}$ = 1.36 (at 589.3 nm and 25°C). The index-matching oil can be readily replaced by a low-index polymer cladding.

## S2. OPTICAL WIRE BOARD COUPLING INTERFACES

The optical wire board (OWB) contains a multitude of coupling sites realized by tapered silicon nanowire waveguides onto which the OWB connections can be printed [1]. A schematic of the coupling site with a printed polymer coupling structure is shown in Supplementary Fig. S1. The taper of the OWB connection (red) starts with an initial cross section of 2.0 µm × 1.8 µm, which is linearly tapered down to a width of 0.76 µm and a height of 0.46 µm. The length of the tapered polymer section as well as of the silicon taper (dark blue) is 60 µm. The Si taper starts with a measured tip width of 0.16 µm and is then converted linearly to the width of 0.5 µm of the Si nanowire waveguide. The Si nanowire has a height of 0.22 µm. The pitch of neighboring Si nanowires is 10 µm, the distance between opposing Si taper tips on the OWB is 150 µm, and the thickness of the underlying buried silicon dioxide (SiO2, light blue) is 2 µm. The OWB coupling interfaces on the chips used in our experiments are non-ideal, which is the reason for the additional coupling losses. Supplementary Fig. S2 shows a SEM image of an OWB under an angle of 54°. Polymer tapers are printed on some of the Si waveguide tips. The inset shows a cut through the Si inverse taper tip obtained by focused ion-beam (FIB) milling. The FIB cut shows that the Si tapers are over-etched during fabrication of the silicon photonics (SiP) chip, which leads to increased insertion loss. Note that this problem is not fundamental and can be overcome by optimizing the SiP fabrication processes, thus permitting losses of 1 dB or less for an OWB connection with two coupling interfaces [2].

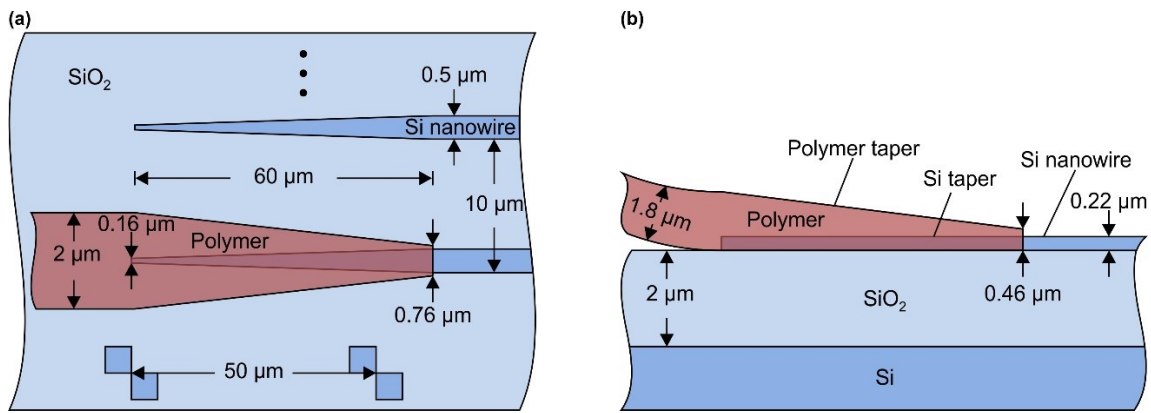

**Fig. S1.** Schematic of the transition of the 3D-printed polymer waveguide to the Si nanowire waveguide. The figures show the top (a) and the side (b) view of the double-taper structure. The polymer taper (red) starts a with an initial cross section of 2.0 µm × 1.8 µm of the optical wire board connection and is linearly tapered down to a width of 0.76 µm and a height of 0.46 µm. The length of the tapered polymer section as well as of the silicon taper is 60 µm. The measured tip width of the Si inverse taper is 0.16 µm and is converted linearly to the width of 0.5 µm of the Si nanowire waveguide. The Si nanowire has a height of 0.22 µm. The pitch of neighboring Si nanowires is 10 µm. The thickness of the underlying buried silicon dioxide (SiO2) amounts to 2 µm.

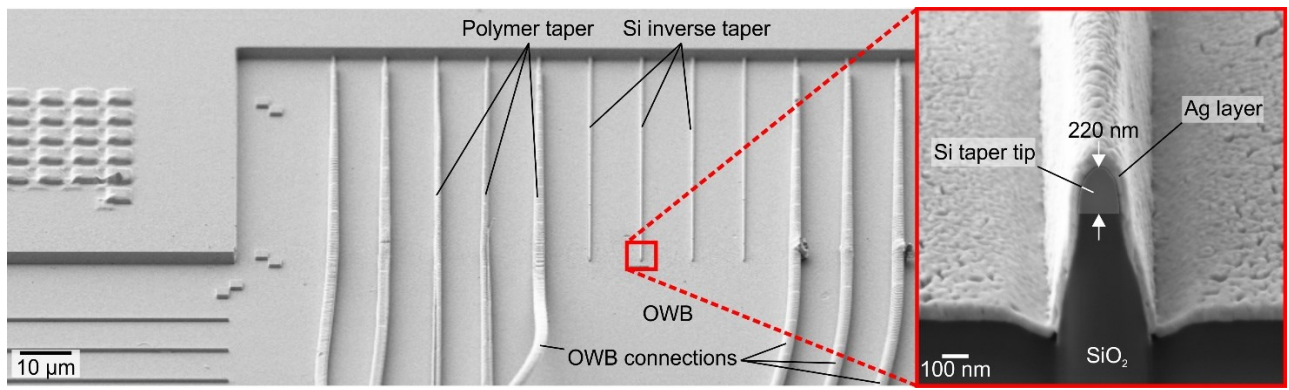

**Fig. S2.** Focused ion beam cut through coupling interfaces of the optical wire board (OWB): Scanning-electron microscope (SEM) image of the coupling interfaces of the OWB. The viewing angle is 54°. Polymer waveguides are printed on some of the inverse Si tapers. Note that this device was rather for test purposes than for real application. The inset shows a focused-ion-beam (FIB) cut through the tip of the Si inverse taper. The sample was sputter-coated with a thin Ag layer to increase conductivity for better SEM imaging. The FIB cut reveals that the Si tapers are strongly over-etched during fabrication process, which leads to increased insertion losses of the OWB coupling interface in our experiments. The specified height of the Si waveguide core is 220 nm.

## S3. INSERTION LOSS MEASUREMENTS

The insertion losses (IL) of the OWB connections are obtained by measuring the transmission of the whole device and by taking into account the grating-coupler (GC) losses as well as the IL of the connected building blocks, e.g., the modulator, which were derived from measurements of on-chip reference structures. For all transmission measurements, we couple light from a tunable laser source (TLS) with an output power of 0 dBm into the chip and detect the optical output power. In all measurements, the chips and the fiber tips were immersed into an index-matching oil, which mimics a low refractive-index overcladding of the 3D-printed OWB connections.

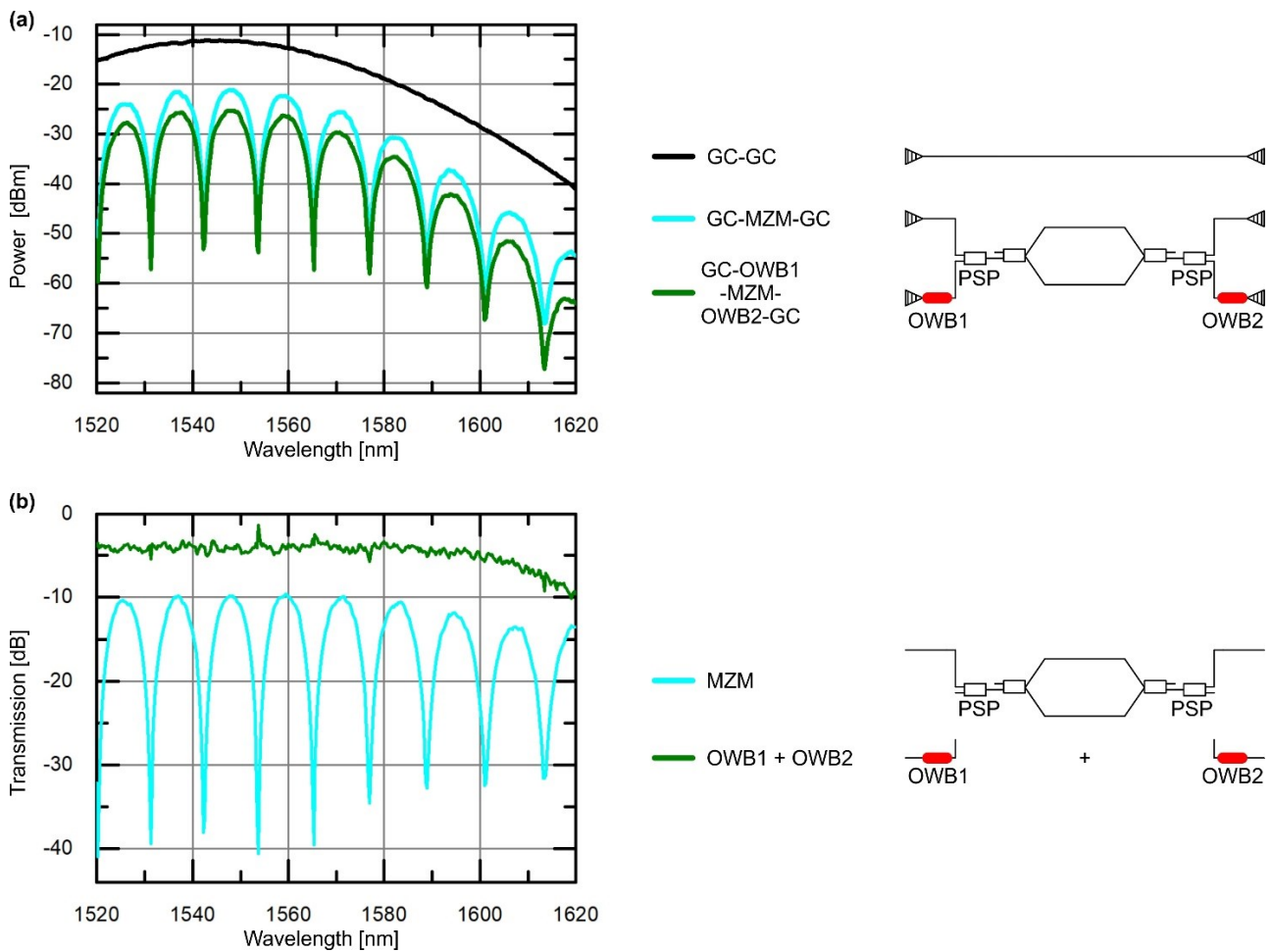

**Fig. S3.** Measurement of OWB insertion loss for the first transceiver implementation (TRx1) as depicted in Fig. 2(a) of the primary document. **(a)** The graph shows the optical transmission obtained for different sequences of on-chip devices, depicted on the right-hand side. A straight on-chip waveguide with two adjacent grating couplers (GC) is used as a reference, black line. To isolate the loss of the on-chip MZM, the device is equipped with a pair of power splitters (PSP), which allow to measure the transmission either directly through a pair of GC, cyan line, or through a pair of GC and a pair of OWB connections (OWB1 and OWB2), green line. The difference of both reveals the combined insertion loss (IL) of the two OWB connections. The MZM was defined to have different arm lengths thus permitting adjustment of the operating point by tuning the wavelength. This leads to pronounced transmission fringes. **(b)** On-chip transmission spectrum of the MZM (cyan) and of the corresponding OWB connections OWB1 and OWB2 (green) as obtained from the measurements described in Subfigure (a). The overall IL of the concatenated OWB connections amounts up to 4.0 dB. The slight increase towards higher wavelengths is attributed to measurement uncertainties caused by high GC losses. The losses of the OWB connections may be reduced further by optimized fabrication, see Supplementary Section S2. The combination of the MZM and the two PSP has an IL of about 10 dB, measured at the top the transmission fringes.

### A. TRx1 and TRx2

Supplementary Fig. S3 and Supplementary Fig. S4 show the results of the reference measurements as well as the transmission measurement of the first transceiver implementation (TRx1). The configuration of TRx1 is described in Fig. 2(a) of the primary document. The transmitter consists of a SiP p-n depletion-type Mach-Zehnder Modulator (MZM) with its input and output waveguides connected to grating couplers (GC) via the OWB. As reference for all subsequent measurements, we quantify the transmission of a pair of GC connected by a straight on-chip waveguide (GC-GC), black line in Supplementary Fig. S3. To allow for separation of the IL of the MZM and OWB connection, we have included two additional power splitters (PSP) into the on-chip waveguides to and from the MZM. We use these PSP to record the transmission of the MZM and the adjacent power splitters via a first set of GC (GC-MZM-GC), cyan line, as well as the transmission of the OWB1-MZM-OWB2 sequence and the PSP via a second set of GC (GC-OWB1-MZM-OWB2-GC), green line.

Note that the MZM was defined to have different arm lengths to permit adjustment of the operating point by tuning the wavelength. This leads to the transmission fringes seen in Supplementary Fig. S3. Note also that the taps associated with the first set of GC are not shown in Fig. 1(b) of the primary document for simplicity. From the measurements, we can derive the on-chip transmission loss spectra of the concatenated OWB connections as well as of the MZM and the adjacent PSP, see green and cyan line in Supplementary Fig. S4, respectively. The combination of the MZM together with the PSP has a total on-chip IL of about 10 dB, measured at the maximum of the transmission fringes, and the overall IL of the concatenated OWB connections amounts up to 4.0 dB. To compensate for the fibre-chip coupling losses and the IL of the MZM, we use an EDFA in the data transmission experiments as shown in Fig. 3 of the primary document. To measure the IL of the OWB connection of the receiver (Rx) part of TRx1, we adjust the power in the optical fibre to a fixed value and then measure the associated photocurrent obtained from a pair of on-chip photodetectors (PD) – one which is directly connected to a GC, and another one, which is connected to the GC via the OWB. The IL of the OWB connection is then calculated by considering the ratio of the two photocurrents and amounts to 0.9 dB.

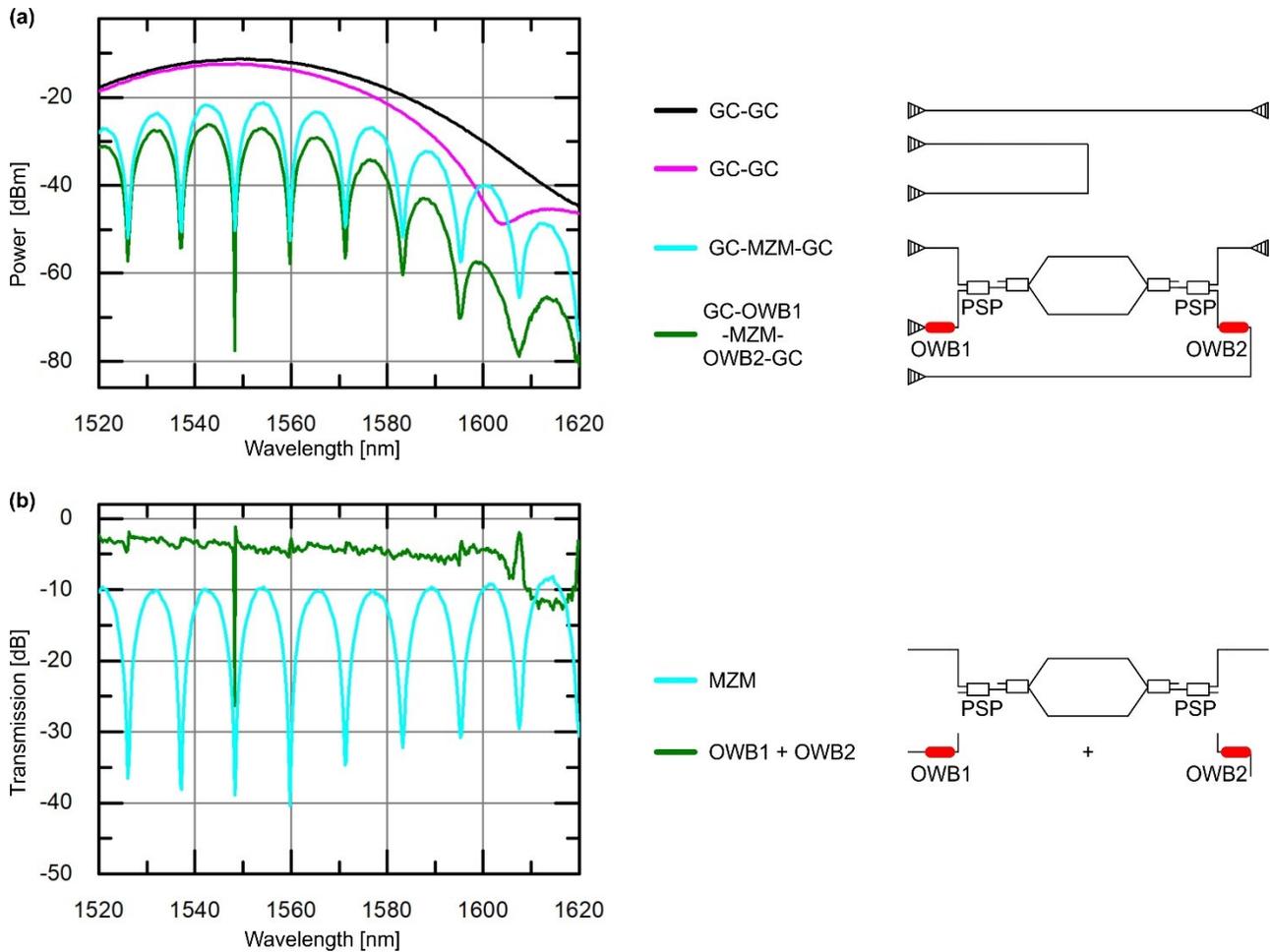

**Fig. S4.** Measurement of OWB insertion loss for the second transceiver implementation (TRx2) depicted in Fig. 2(b) of the primary document. **(a)** The graphs show again the optical transmission obtained for different sequences of on-chip devices, depicted on the right-hand side. As references we use a straight on-chip waveguide with two adjacent GC, black line, as well as an on-chip waveguide with GC, configured for in and out-coupling of light with the same fibre array, magenta line. In the same manner as described for TRx1 we measure the transmission of the MZM including the adjacent PSP either directly through a pair of GC, cyan line, or through a pair of GC and a pair of OWB connections, green line. The combined insertion loss (IL) of the two OWB connections (OWB1 and OWB2) is obtained by taking the difference of the cyan and the green line and by accounting for the corresponding GC-GC references. **(b)** On-chip transmission spectrum of the MZM and the adjacent PSP (cyan) as well as of the corresponding OWB connections OWB1 and OWB2 (green) as obtained from the measurements described in Subfigure (a). The overall IL of the concatenated OWB connections amounts to approximately 4.0 dB. The slight increase towards higher wavelengths and the outlier around 1550 nm are attributed to thermal drift of the MZM transmission spectrum and to measurement uncertainties caused by high GC losses. The losses of the OWB connection may be reduced further by optimized fabrication, see Supplementary Section S2. The combination of the MZM and the two PSP has an IL of about 10 dB, measured at the top the transmission fringe.

The implementation of TRx2 is depicted in Fig. 2b of the primary document, and the transmission measurements are shown in Supplementary Fig. S4(a). The strategy of extracting the IL of the OWB connections is similar to the one described for TRx1. Note that for TRx2, fibre arrays rather than individual fibres are used for in-coupling and out-coupling of light and a separate reference structure is thus needed to quantify the associated coupling losses. We again use plain on-chip waveguides connected to a pair of grating couplers for reference measurements with two independent single mode fibres (SMF), black line in Supplementary Fig. S4(a), and with two fibres of a common array, magenta line. The transmission of the MZM and the adjacent PSP is measured with two independent SMF, cyan line in Supplementary Fig. S4(a), whereas the transmission with additional OWB connections is obtained by using a fibre array, green line. From this data, we extract the combined IL of the MZM and PSP, cyan line in Supplementary Fig. S4(b), as well as the losses of the two OWB connections OWB1 and OWB2, green line. The overall IL of the concatenated OWB connections again amounts up to 4.0 dB, and the combined IL of the MZM and the PSP is about 10 dB. The IL of the OWB connection to the Rx photodiode is measured in the same way as for TRx1 and amounts to 1.8 dB.

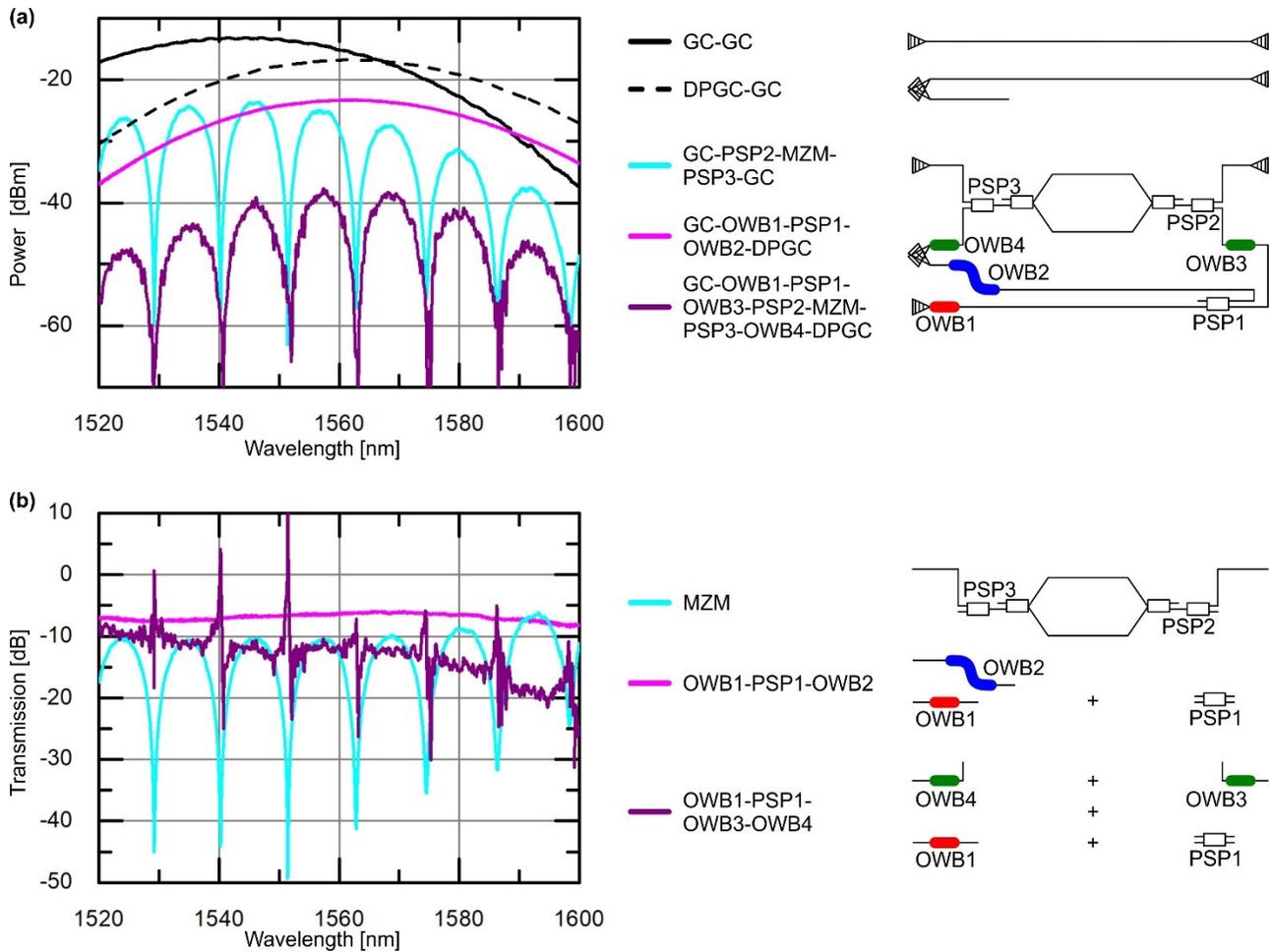

**Fig. S5.** Insertion loss measurements for the self-homodyne transmitter implementation as depicted in Fig. 2(c) of the primary document. **(a)** The graphs show again the optical transmission spectra obtained for different configurations of on-chip devices, as depicted on the right-hand side. We measured an on-chip reference for the grating couplers (GC) with two single mode fibres (SMF), black line, a reference for the dual-polarization grating coupler (DPGC) connected to a standard single-polarization GC, black dashed line, and a reference of the Mach-Zehnder modulator (MZM), cyan line. The magenta line corresponds to the transmission spectrum of the LO path (GC-OWB1-PSP1-OWB2-DPGC), whereas the purple line represents the signal path (GC-OWB1-PSP1-OWB3-PSP2-MZM-PSP3-OWB4-DPGC). **(b)** The transmission spectrum of the MZM is depicted in cyan, whereas the transmission of the on-chip LO path (OWB1-PSP1-OWB2) and the on-chip signal path (OWB1-PSP1-OWB3-OWB4) are depicted in magenta and purple. Taking into account the additional 3 dB of power splitter PSP1, we obtain an overall insertion loss of 3.3 dB measured for the concatenation of OWB1 and OWB2 connections along the LO path and of 7.6 dB for the concatenation of OWB1, OWB3, and OWB4 along the signal path. The irregularities in the transmission spectrum of the on-chip signal path are caused by polarization crosstalk in the DPGC in combination with the fact that the on-chip LO path features a relatively strong transmission compared to the on-chip signal path.

## B. Self-homodyne transmitter

The optical losses of the OWB connections of the self-homodyne transmitter implementation are obtained from the results of reference measurements shown in Supplementary Fig. S5(a). As before, we measure two reference GC on the chip, connected by a SiP waveguide, black solid line, as well as a reference of a dual-polarization grating coupler (DPGC) connected to standard single-polarization GC, black dashed line. For the transmitter, we first measure the LO path, containing the input GC, two OWB connections (OWB1, OWB2), a power splitter (PSP1), and a DPGC (GC-OWB1-PSP1-OWB2-DPGC), magenta line. For the MZM and the adjacent PSP (PSP2, PSP3), we take two measurements – one through a pair of GC (GC-PSP2-MZM-PSP3-GC), and one through the signal path containing the input GC, three OWB connections (OWB1, OWB3, OWB4), three power splitters (PSP1, PSP2, PSP3), and the MZM (GC-OWB1-PSP1-OWB3-PSP2-MZM-PSP3-OWB4-DPGC), purple line. Note that the signal and LO path share one OWB (OWB1) as well as one power splitter (PSP1). Using the reference measurements, we can then isolate the transmission of the MZM, the transmission of the on-chip LO path (OWB1-PSP1-OWB2), and the transmission of the on-chip signal path (OWB1-PSP1-OWB3-OWB4), see Supplementary Fig. S5(b). The transmission of the MZM including the adjacent PSP (PSP2, PSP3) is drawn in cyan and is about 10 dB, again similar to the results obtained for TRx1 and TRx2. The transmission through the on-chip LO path and the on-chip signal path are depicted in magenta and purple, respectively. The irregularities in the transmission spectrum of the on-chip signal path (purple line) are caused by polarization crosstalk in the DPGC, originating from the fact that a relatively strong signal is coupled into the lower port of the DPGC via the on-chip LO path, which leads to a noticeable cross-talk signal superimposed onto the rather weak signal transmitted by the lossy on-chip signal path. This effect was partially compensated by estimating the spurious signal and by subtracting it from the measured transmission spectrum that is depicted as a purple line in Supplementary Fig. S5(a). Still, some distortions remain, which are most pronounced close to the transmission dips of the MZM, where the spurious polarization cross-talk dominates over the transmission of the on-chip signal path. In addition, some outliers can be attributed to thermal drift of the MZM transmission spectrum. If we account for an additional IL of 3 dB due to PSP1, we can estimate an overall loss of approximately 3.3 dB for the concatenated OWB connections along the LO path (OWB1, OWB2) and approximately of 7.6 dB for the OWB connections along the signal path (OWB1, OWB3, OWB4). This leads to an average IL of approximately 1.65 dB per OWB connection in the LO path and of approximately 2.53 dB per OWB connection in the signal path. Note that, due to the lack of appropriate reference structures, all reference measurements for the GC and DPGC were performed with a pair of SMF rather than with a fibre array, as used for measuring the IL of the on-chip transmitter structure. This leads to a slight overestimation of the OWB IL.

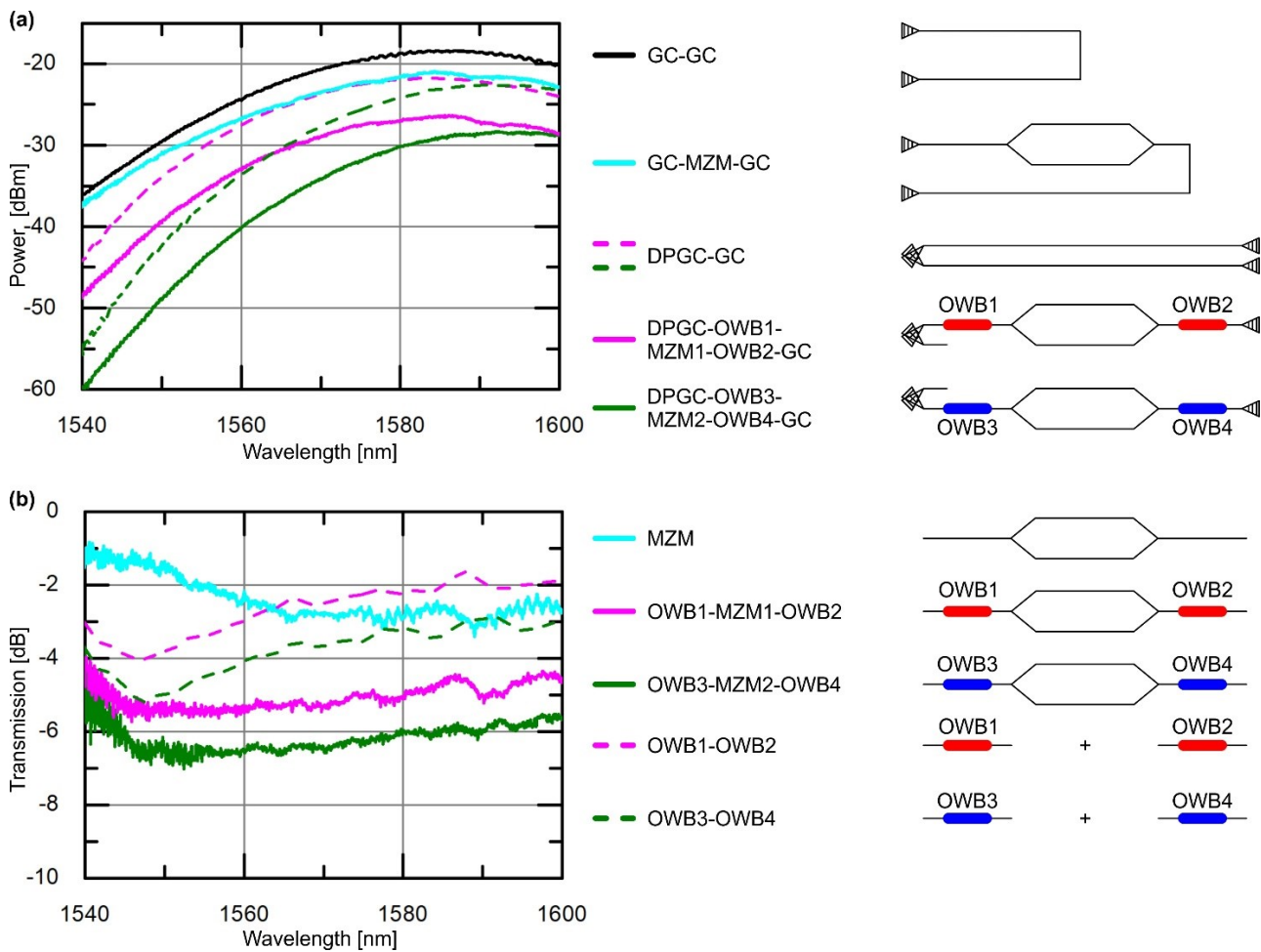

**Fig. S6.** Insertion loss measurements for the dual-polarization transmitter implementation as depicted in Fig. 2(d) of the primary document. **(a)** The graphs show again the optical transmission spectra obtained for different configurations of on-chip devices, as depicted on the right-hand side. We measured an on-chip reference for the grating couplers (GC) used for fibre-array coupling, black line, and a reference of a Mach-Zehnder modulator (MZM) which is nominally identical to the ones used in the transmitter, cyan line. References for the two outputs of dual-polarization grating coupler (DPGC) connected to a standard single-polarization GC are depicted as dashed magenta and green lines. Note that, due to imperfections of the DPGC, the maxima in transmission of both outputs are located at slightly different wavelengths. The magenta (green) line shows the transmission of the first (second) transmitter path Tx1 (Tx2). **(b)** The transmission spectrum of the MZM is depicted in cyan, and the transmission of the OWB connections of the transmitter paths Tx1 and Tx2 including the MZM are depicted in magenta and green. This reveals the transmission spectra of the OWB connections in each signal path, which are depicted as dashed magenta and green lines. The insertion loss of the concatenated OWB connections OWB1-OWB2 (OWB3-OWB4) amounts to approximately 1.7 dB (3.1 dB) at a wavelength of 1587 nm, where the GC reach their maximum efficiency. The loss increases slightly towards smaller wavelengths, which is attributed to imperfect reproducibility of the grating coupler losses across the various reference measurements and structures.

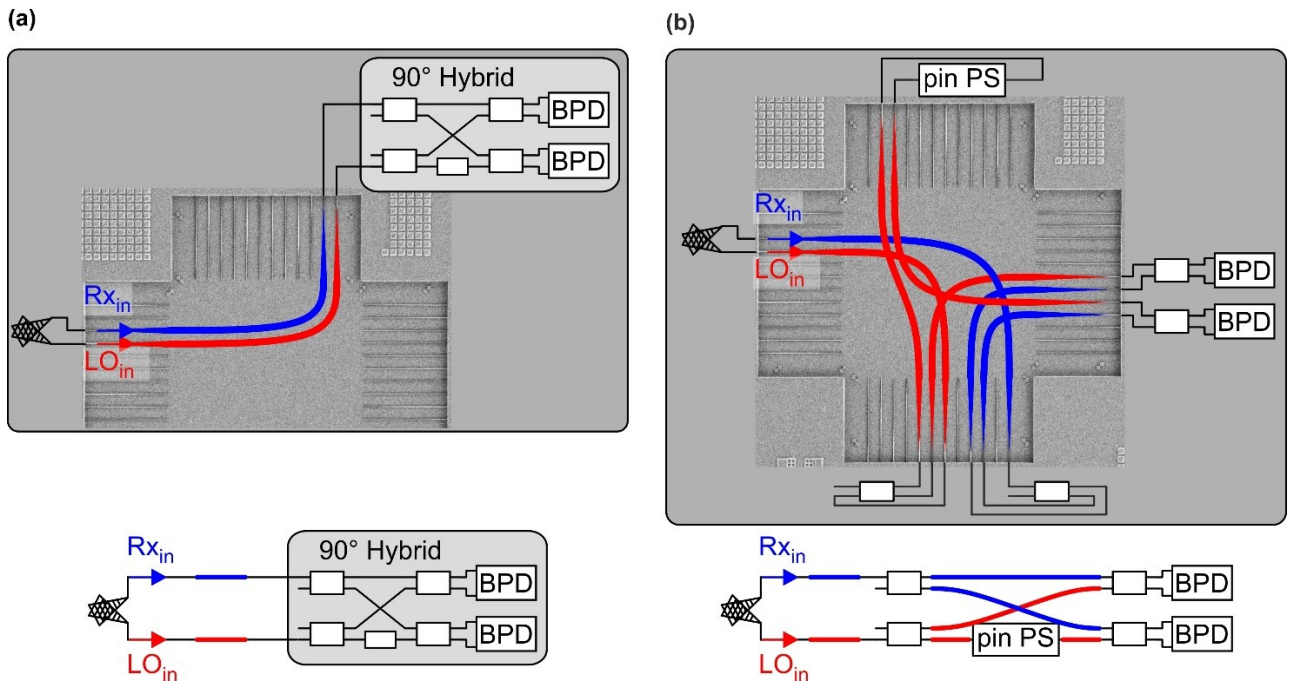

**Fig. S7.** Schematic drawings of two self-homodyne receiver implementations realized with the HC-PIC. **(a)** Self-homodyne receiver comprising a dual-polarization grating coupler (DPGC) connected to a coherent receiver, which consists of a 90° optical hybrid and two pairs of balanced photo detectors (BPD). **(b)** The 90° optical hybrid is realized by concatenating appropriate building blocks via OWB connections. In both configurations, a polarization-maintaining input fibre is needed to transmit the LO and the data signal on two orthogonal linear polarization states, which are separated by the DPGC. The device can be extended to work with a standard single-mode fibre (SMF) by including an on-chip polarization controller [3] between the DPGC and the optical hybrid.

### C. Dual-polarization transmitter

The results of the reference measurements for the dual-polarization transmitter are shown in Fig. S6. We measure an on-chip reference for the GC for fibre-array coupling, black line, and an MZM which is nominally identical to the ones used in the dual-polarization transmitter, cyan line. As a reference for the DPGC, we couple light into the device from a standard single-mode fibre (SMF) and adjust the polarization for maximum transmission to either of the two connected single-polarization GC, thus leading to dashed magenta and green lines in Supplementary Fig. S6(a). Note that the maximum transmission for the polarizations occurs at slightly different wavelengths and at slightly different positions of the SMF with respect to the DPGC, indicating that the device design leaves room for improvement. Moreover, we measure the transmission spectrum of the first (second) transmitter path Tx1 (Tx2), leading to the solid magenta (green) lines in Fig. S6(a). From these measurements, we extract the on-chip transmission spectra of the reference MZM, cyan line in Fig. S6(b), and of the two on-chip signal paths, each containing an MZM as well as a pair of OWB connections (OWB1-MZM1-OWB2, OWB3-MZM2-OWB4), solid magenta and green lines in Fig. S6(b). This reveals the transmission spectra of the OWB connections in each signal path, which are depicted as dashed magenta and green lines in Fig. S6(b). At a wavelength of 1587 nm, which corresponds to the maximum GC efficiency, we obtain IL of 1.7 dB (3.1 dB) for the concatenated OWB connections OWB1-OWB2 (OWB3-OWB4). These values increase slightly towards smaller wavelengths, which we attribute to imperfect reproducibility of the grating coupler losses across the various reference measurements and structures.

### S4. EXPERIMENTAL DEMONSTRATION

In the transmission experiments, we used a benchtop-type arbitrary-waveform generator (AWG; Keysight M8196A) to synthesize the electrical drive signals for the modulators. These signals are boosted by RF amplifiers and then fed to the chip through microwave probes in ground-signal-signal-ground (G-S-S-G) configuration. For the experiment depicted in Fig. 3 of the primary document, we used an external photodiode (Finisar XPDV2320R) with an electrical 3 dB-bandwidth of 50 GHz as a reference receiver. For the self-homodyne experiment shown in Fig. 4 of the primary document, we used an optical modulation analyzer (Keysight N4391) along with high-speed oscilloscopes (Keysight DSOX93204A) to receive the BPSK and 4-ASK signals. In all experiments, amplified spontaneous emission (ASE) noise of EDFA was suppressed by a tunable bandpass filter with a full width at half maximum (FWHM) of 0.6 nm. For the IM-DD measurements shown in Fig. 3 of the primary document, the electrical drive signals were pre-distorted to compensate for the measured frequency response of the respective MZM. The received waveforms were analyzed by signal processing routines implemented in Python, which comprise filtering, clock recovery, equalization and resampling. For each experiment, we analyzed $3.9 \times 10^7$ transmitted bits. For OOK, no errors could be found within the signal sequence such that we can only specify an upper bound of $2.6 \times 10^{-8}$ for the corresponding bit error ratio (BER). For the self-homodyne transmission experiment, Fig. 4 of the primary document, signal processing and BER extraction is done using the Keysight Vector Signal Analysis (VSA) software, comprising clock recovery, resampling, adaptive equalization, and carrier recovery for demodulation of the received signal. Here we analyzed $10^6$ bits for each experiment. For BPSK signaling, no errors could be found within the analyzed signal sequences such that we can only specify an upper bound of $1.0 \times 10^{-6}$ for the corresponding BER.

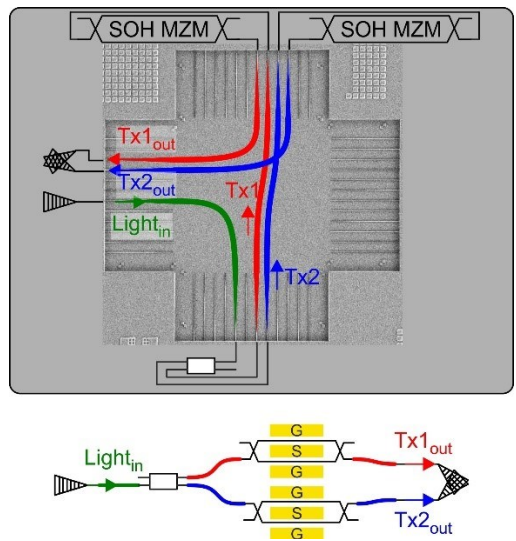

**Fig. S8.** Alternative implementation of a dual-polarization transmitter. The device comprises an additional power splitter to distribute light from a single optical source to both modulators. In the depicted configuration, fast silicon-organic hybrid modulators (SOH MZM) could be used for encoding of data onto the two polarizations of the light signal. For the PIC used in our experiments, this turned out to be dysfunctional due to fabrication errors of the SOH devices. This is not a fundamental problem [4] - in future chip designs, the Mach-Zehnder modulators (MZM) may hence be replaced by high-performance IQ modulators to fully exploit the benefits of polarization-diverse coherent transmission.

## S5. FURTHER FUNCTIONALITIES OF THE GENERIC PIC

The transceiver demonstrations described in the primary document only represent a few examples of the possible PIC configurations that can be realized by appropriate concatenation of the available building blocks. Further configurations comprise, e.g., self-homodyne receivers or a dual-polarization transmitter that is fed by a single light source, see discussion of Fig. 2(c) and (d) in the primary document. Supplementary Fig. S7 shows two possible configurations for realizing self-homodyne receivers. The configuration shown in Supplementary Fig. S7 (a) relies on an on-chip coherent receiver, i.e., a 90° optical hybrid followed by two pairs of balanced photo detectors (BPD), connected to a DPGC via the OWB. In the second configuration, Fig. S7 (b), a second set of BPD is used, and the 90° optical hybrid is realized by a concatenation of power splitters and an adjustable phase shifter (pin PS). In both configurations, a polarization-maintaining input fibre is needed to transmit the LO and the data signal on two orthogonal linear polarization states which are separated by the DPGC. The device can be extended to work with a standard single-mode fibre (SMF) by including an on-chip polarization controller [3] between the DPGC and the 90° optical hybrid.

Supplementary Fig. S8 shows an alternative configuration for realizing a dual-polarization transmitter. In contrast to the configuration described in Fig. 2(d) of the primary document, the device depicted here relies on an additional power splitter to distribute light from a single optical source to both modulators. Moreover, this configuration relies on fast silicon-organic hybrid (SOH) Mach-Zehnder modulators (SOH MZM), and not on the comparatively slow pin MZM that were used for the implementation described in the main manuscript. Note that, for the PIC used in our experiments, the configuration shown in Supplementary Fig. S7 turned out to be dysfunctional due to fabrication errors of the SOH devices. This is not a fundamental problem – other fabrication runs led to SOH devices that could be combined with 3D-printed photonic wire bonds [4] and that should hence be compatible with the HC-PIC concept. In future chip designs, the MZM can be replaced by IQ modulators to fully exploit the benefits of polarization-diverse coherent transmission.